# Sub-picosecond molecular dynamics modulate the energies of spin states for hundreds of nanoseconds

Lorena E. Rosaleny,[1,2] Kirill Zinovjev,[1] Iñaki Tuñón,[1] Alejandro Gaita-Ariño*[2]

[1] Departament de Química Física, Universitat de València, 46100 Burjassot, Spain

[2] Instituto de Ciencia Molecular, Universitat de València, 46980 Paterna, Spain

Molecular vibrations are increasingly seen as a key factor for spin dynamics in single-ion magnets and molecular spin qubits. Herein we show how an inexpensive combination of molecular dynamics calculations and a crystal field analysis can be employed to obtain a dynamical picture of the crystal field splitting. We present the time evolution during 500ns of the spin energy levels in a terbium complex. Our calculations evidence an amplitude of up to tens of cm$^{-1}$ for the oscillations of the spin energy levels in the fs time scale. Crucially, we also see that the oscillations average out and practically disappear at longer time scales, thus ruling out the modulation of spin energy levels by phonons of long wavelength λ (up to λ = 40 Å). Our results are compatible with the common approximation of focusing on local vibrations, but at the same time highlight the risks of assuming that the spin energy levels are time- and temperature-independent.

Molecular spin qubits and single ion magnets (SIMs) have recently achieved a jump upward in terms of their performances, with spin qubits presenting phase memory times of almost 1 ms[1] and magnetic hystereses of SIMs up to 60 K.[2] The latter result was unexpected, with hystereses of single-molecule magnets having experienced a rather slow progress since its beginning more than two decades ago.[3] During this time, much effort has been put into the theoretical description of the crystal field Hamiltonians of magnetic molecules,[4] which has allowed designing molecules with effective barriers of up to 1000 cm$^{-1}$, but hystereses typically below 10 K.

Only recently has the modelling of the coupling between spins states and vibrations been recognized as a key factor for the slow relaxation of magnetisation at higher temperatures.[2,5,6] Still, this has so far been done under a time-independent crystal field Hamiltonian determined by ligands in static positions. However, considering that an irreversible energy exchange takes place between the spin subsystem and the thermal bath via the vibrational subsystem, it would be interesting to obtain information about the real-time interaction between the spin dynamics and the dynamics of the molecular structure, at all relevant time scales.

Let us start by pointing out that the time scale of spin dynamics worth studying in magnetic molecules spans many orders of magnitude: between $10^{-12}$ s and $10^{-3}$ s, and beyond. Consider SIMs: one often describes their magnetic relaxation at high temperature with the Arrhenius equation, with a wide range of characteristic times $10^{-11}$ s > $\tau_0$ > $10^{-6}$ s being reported in the literature for the pre-exponential factor.

In the case of qubits, the shortest relevant time limit is the operational time necessary to invert the value of a spin qubit. This is typically achieved with a pulse of microwave radiation of the order of $10^{-8}$ s. The longest interesting times are the relaxation times $T_1$, $T_2$, $T_m$. Phase memory times $T_m$ of an ensemble of spins are measurable on standard pulse Electronic Paramagnetic Resonance (EPR) between $T_m = 10^{-7}$ s and $T_m = 10^{-3}$ s.[1] Here note that individual rare events that are responsible for relaxation might be much faster, but their nature is not well understood. Experiments based on UV-vis photons deserve a comment: optical processes are much faster but they also involve the spin states which are to be used as qubits. In the context of lanthanide complexes, spectroscopy is currently only used as a characterization technique, but other spin qubits, most notably nitrogen-vacancy centers, do rely on optical transitions, so optical processes also have potential importance for the future of this field. In this case, the relevant time scale for excitations is in the order of $10^{-15}$ s.

In the present work we implemented the first *in silico* exploration of the explicit time-dependent dynamics of spin energy levels due to thermal molecular movements. As a model system we chose a family of peptides known as Lanthanide Binding Tags (LBTs), developed in the field of Biochemistry for the study of protein structure and dynamics,[7] since some of us recently showed that LBTs are promising both as spin tags to open a new playground for molecular spin qubits,[8,9] and as hardware in molecular spintronics to prepare single-component spin valves.[10] As a magnetic ion we chose $Tb^{3+}$ because of the interesting spectroscopic properties of TbLBT[7] and because, being a non-Kramers ion, $Tb^{3+}$ is a good model system to study the effects of geometrical distorsions on tunneling splitting.

Herein we combined two inexpensive computational tools, namely molecular dynamics (MD) and a crystal field modelling based on effective point charges. The LBT structure was obtained from Protein Data Bank (PDB code 1TJB).[11] The MD parameters for the $Tb^{3+}$ ion were obtained from Quiao *et al.*[12] The MD simulations were performed in a mixture of water/glycerol reproducing the experimental samples for EPR experiments.[9] The NPT ensemble and pmemd program from the Amber2017 package were employed. Simulations were carried out at 300 K, 30 K and 3 K. The coordinates at each snapshot were then used as an input for SIMPRE, a software package that employs an effective point-charge model to inexpensively estimate the spin energy levels in lanthanide complexes.[13] Combined with the Radial Effective Charge model,[14,15] this approach has been shown to provide a level of accuracy comparable to ab initio methods.[4] Extensive methodological details can be found as S.I., including details of the system preparation and the MD protocol.

We ran the simulations for 500 ns. A sparse sampling of the 13 spin energy levels of the $J$ = 6 ground multiplet serves to estimate the energy dispersion in the spin energy levels when probed in the femtosecond timescale at 30 K (Fig. 1). This dispersion due to vibrations is of the order of 10% of said crystal field energies (or about ±50 $cm^{-1}$) for the highest levels, and about 100% of the energy (or about ±10 $cm^{-1}$) for the first excited level, corresponding to the tunneling splitting of the ground doublet. Note that in a sample that is measured in an optical spectrometer, there is an ensemble of molecules with different instantaneous geometries. Assuming that the evolution of a single molecule provides a good sampling of the instantaneous state of an ensemble of molecules (e.g. is ergodic). The calculated dispersion is enough to impact the spectroscopic linewidths and thus to determine the practical limit of the precision in the determination of these energies and the underlying crystal field parameters.

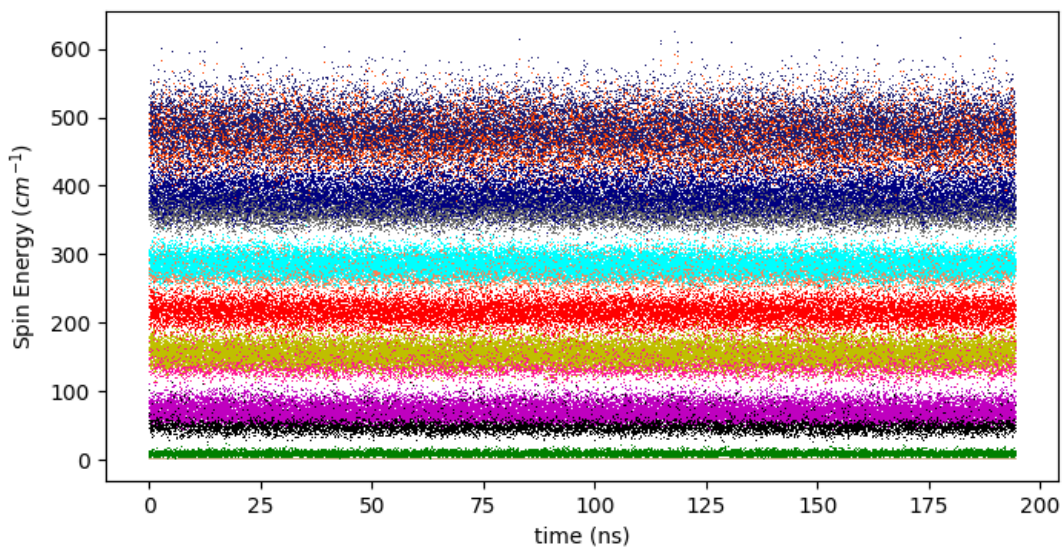

**Figure 1:** Spin energy levels in the ground $J$ multiplet of TbLBT during a ~500 ns trajectory at 30 K as calculated by AMBER+SIMPRE, instantaneous energies.

When calculations are performed both for 30 K and 3 K, and represented with a time lapse of $2 \cdot 10^{-15}$ s, it is possible to recognize the underlying smooth structure (Fig. 2) of the energy modulation. The large amplitude of the energy variability means that the vibrations that take place in the time scale of the tens of femtoseconds are strongly coupled to the spin energy levels. It is also possible to observe that the temperature greatly affects both the modulation amplitude of the first excited spin energy level and its average value. Moreover, the autocorrelation of the first excited spin energy level calculated over the full period of 500 ns at 30 K and 3 K allows to recognize the same patterns at both temperatures, which means that they are presumably related to the same molecular vibrations. These patterns survive up to times of about 2 ps and then quickly die out, althought in fact most of the decay of the autocorrelation function takes place in the first ~50 fs. A Fourier transform of these data results in a rough determination of the underlying frequencies, with the main one being at ~580 $cm^{-1}$ and two secondary ones appearing at lower frequencies (see Supporting Information section S6).

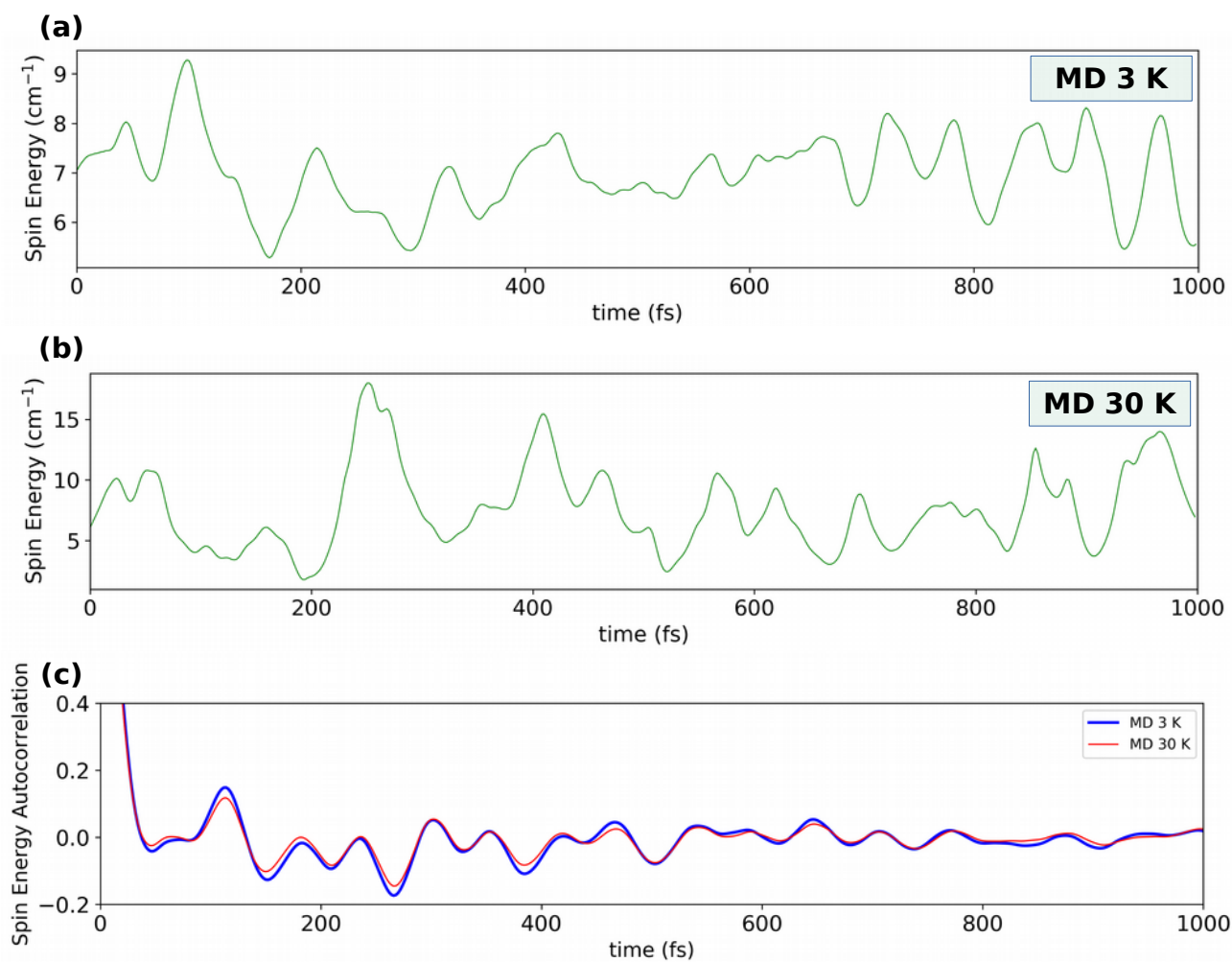

**Figure 2:** Spin energies for the first excited state during two sample 1000 fs trajectories at (a) 3 K and (b) 30 K in a total of 500 steps in each case, with a separation of 2 fs. (c): Superimposed plots for the autocorrelation functions at the two temperatures.

To study non-periodical effects at longer time scales, we represented the effect of the coordination geometry during the MD trajectory on the first excited spin level averaging at longer time scales, from the picosecond to the tens of nanoseconds (Fig. 3). We found a reduction in the energy dispersion that apparently follows a square root law, where a period of time a hundred times longer translates into one tenth of the noise amplitude. In other words, the energy dispersion is monotonically reduced when averaging at longer times, meaning there is no significant modulation that is intrinsic of longer time scales. Instead, the sub-picosecond molecular dynamics are the ones that modulate the energies of spin states up to the sub-microsecond times. The temperature also can have an effect, since increasing the thermal energy by heating the system up from 3 K to 30 K causes the system to increase its noise amplitude by at least a factor 2. We note that in the calculated trajectories there is a single one-time event that is associated to the nanosecond-scale dynamics at 30 K which causes a significant modulation in the energy of the spin states.

In conclusion, these results give us a new insight into the central role of local vibrations even at times that are several orders of magnitude above their characteristic period. If this is verified as general for a variety of systems, it points towards an *a posteriori* justification of the common approximation that involves focusing on local vibrations -or, in periodic calculations, Γ-direction phonons- rather than considering the whole phonon phase space. In contrast, this study quantifies the error commited by the common assumptions of considering the spin energy levels as constant at different times and at different temperatures, since the methodology we employ is able to estimate the the modulation amplitude of the spin energy levels with time and the variation of their average values with temperature.

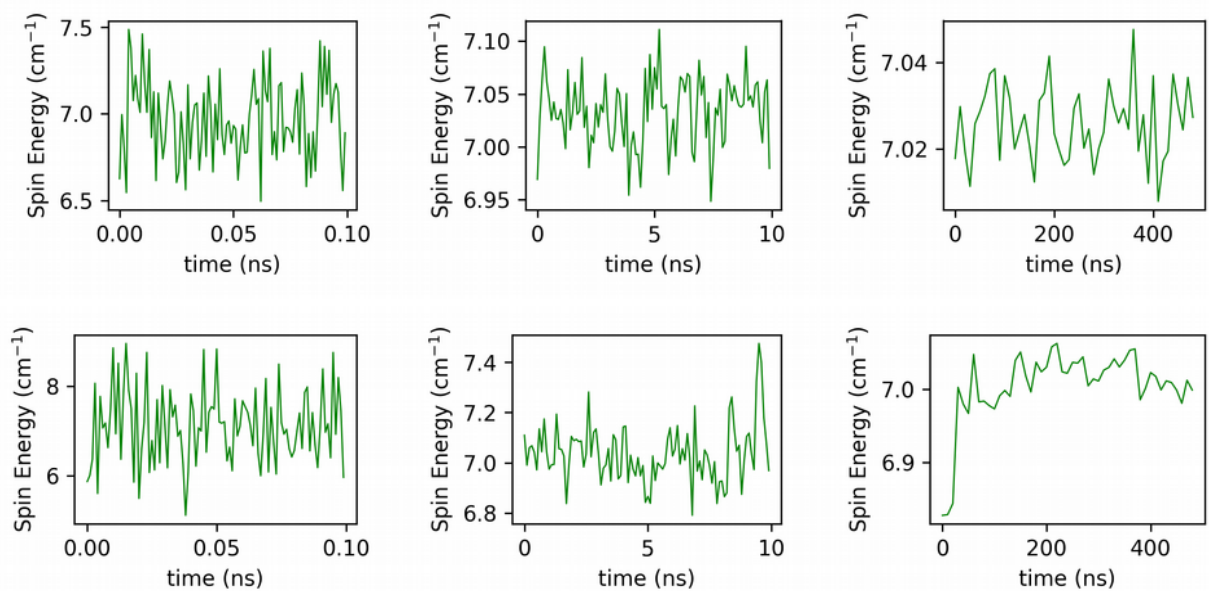

**Figure 3:** From left to right: evolution of the first spin excited state in the ground *J* multiplet of TbLBT, averaging every $10^{-12}$ s, every $10^{-10}$ s or every $10^{-8}$ s (top: 3 K, bottom: 30 K).

**Acknowledgements**
The present work has been funded by the EU (ERC Consolidator Grant 647301 DECRESIM, COST 15128 Molecular Spintronics Project and the European Project SUMO of the call QuantERA), the Spanish MINECO (grant CTQ2015-66223-C2-2-P, grant MAT2017-89528, grant CTQ2017-89993 cofinanced by FEDER and Excellence Unit María de Maeztu MDM-2015-0538), and the Generalitat Valenciana (Prometeo Program of Excellence). AGA acknowledges funding by the MINECO (Ramón y Cajal Program). KZ acknowledges a FPU fellowship from Ministerio de Educación.

Supporting Information

# Sub-picosecond molecular dynamics modulate the energies of spin states for hundreds of nanoseconds

Lorena E. Rosaleny,[1,2] Kirill Zinovjev,[1] Iñaki Tuñón,[1] Alejandro Gaita-Ariño[2]

[1] Departament de Química Física, Universitat de València, 46100 Burjassot, Spain
[2] Instituto de Ciencia Molecular, Universitat de València, 46980 Paterna, Spain

**List of contents**

## S1. System preparation

We extracted one calcium bound LBT molecule from the PDB file (1TJB ID),[1] and stripped all the waters and ions not necessary for the structural integrity. Then the PDB file was edited so that calcium was substituted for terbium. Subsequently input files required for the terbium were created, and this new PDB file was used to create a system consisting of a molecule of TbLBT situated in the middle of a cubic box (side 41.9 Å) of solvent (20% glycerol in water) using the Packmol program.[2] Two $Na^+$ ions were added to ensure the neutrality of the system. The five negative charges in the LBT peptide are countered by the three positive charges belonging to terbium, and the two positive charges contributed by the $Na^+$ ions. The cubic box generated was comprised also 1884 water molecules and 116 glycerol molecules so as to maintain the same water-glycerol ratio as in EPR experiments performed with samples containing similar samples.

## S2. MD protocol

The MD setup was identical for simulations at the three temperatures presented. The temperature and pressure were controlled using Langevin thermostat and Berendsen barostat respectively. The timestep in all cases was 2 fs with bonds involving hydrogens fixed using SHAKE.[3] The following simulation protocol was used: first, the system was heated during 1 ns from 0 K to 300 K, and then was equilibrated at 300 K for 120 ns in the NPT ensemble at a pressure of 1 bar. The production run for this temperture had a duration of 390 ns. The last frame of the simulation at 300 K was used to initiate the production run at 30 K by gradually lowering the temperature from 300 to 30 K during one nanosecond and then equilibrating the system during the next nanosecond at 30 K. The same protocol was used to initiate the production simulation at 3 K. For 30 K and 3 K the production simulation had a duration of 500 ns.

## S3. Spin energy levels calculations: coupling between MD software and spin energy levels software

SIMPRE1.1 is a fortran77 code based on an effective electrostatic model of point charges around a lanthanide magnetic ion to estimate the spin energy levels based on the position (coordinates) of the atoms directly bound to the magnetic ion. In this study case the ion is the lanthanide element $Tb^{3+}$. The package has already been successfully applied to several mononuclear systems with single-molecule magnetic behavior,[4] and, thanks to the parameterization of common ligands as effective charges, it is possible to build upon these results to not only rationalize but also predict the properties of more complex systems. In this context a well-tested geometrical model known as Radial Effective Charge (REC) model was used [5,6], which evaluates the crystal field effect by placing an effective charge along the lanthanide–ligand axes (Fig. S3). This effective charge, which is smaller than the formal charge, and located at a closer distance compared with the crystallographic distance, reproduces the covalency effects.

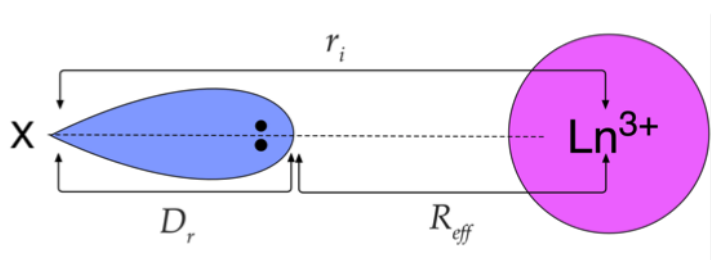

**Figure S3.** Electronic pair of a ligand X oriented towards the nucleus of a trivalent lanthanide cation. The effective charge is located between the lanthanide and the donor atom at $R_{eff} = R_i - D_r$. Extracted with permission from the thesis of J. J. Baldoví, 2016.[4]

The SIMPRE1.1 program allows to estimate the 13 electron spin energy levels of $Tb^{3+}$ in a computationally efficient way. These 13 energy levels arise from the fact that the $Tb^{3+}$ ion has a total angular momentum ($J$) of 6, and as such, the secondary total angular momentum ($M_j$) can take $2J + 1$ values. The input files that SIMPRE1.1 requires are two. One defines the computational parameters and yes/no switches for the calculation (simpre.par). The second input file (simpre.dat) defines the coordinates for the oxygen atoms that act as ligands taking the metal atom as origin of coordinates, contains the effective point charge values for these atoms, and also defines some other output options.

In order to build the simpre.dat file, the coordinates of the potential oxygens involved in the sphere coordination of the $Tb^{3+}$ ion were selected and extracted from the MD run employing a fortran90 script implemented in-house. This script served to communicate with the parallel AMBER pmemd.cuda code (MPI).

This script requires an input file that defines the number of atoms to extract the coordinates from, the number of time steps between two consecutive extractions of coordinates, and the indexes of the relevant atoms to collect the coordinates from. The cartesian coordinates of atoms, from what we can think of as snapshots of the TbLBT molecule every a certain number of time steps, are written in the output file one line for time step. It has to be noted that the spacing between time steps was set in the MD simulations as 2 fs, and this has a consequence on how separated are the snapshots/frames we obtain.

**S3.1. Python script to manage calculations of electronic spin energy levels for MD snapshots**

The analysis of the MD results was performed using a 2.7 python script developed during the present work. This python script uses as input the cartesian coordinates of atoms, and a general simpre.par file, to estimate the spin energy levels for each MD snapshot. Another input file was required to identify the procedence of the distinct oxygens that coordinate the metal, so that effective charges and reduced radius could be assigned before the generation of the simpre.dat file. The effective charges and reduced radius used can be obtained from Rosaleny *et al*., 2018.[7] The obtention of the spin energy levels using the script required the execution in a loop of SIMPRE1.1 for each snapshot.

## S4. Spin Energy Levels at 3 K, 30 K and 300K

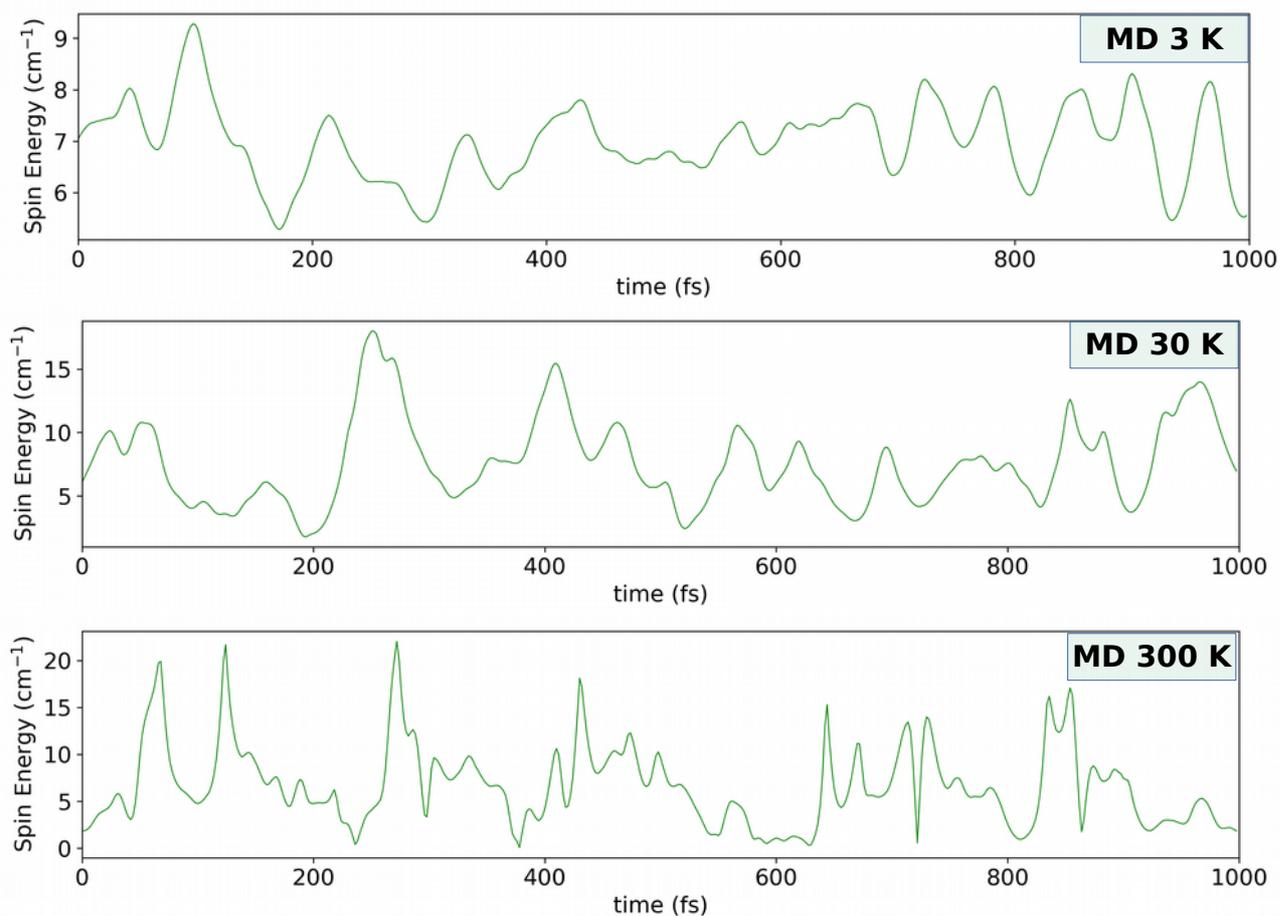

**Figure S4.** Spin energy for the second level for MD at 3 K, 30 K and 300 K. All time steps produced for the depicted MD run are plotted for one 1ps.

# S5. Autocorrelation plots

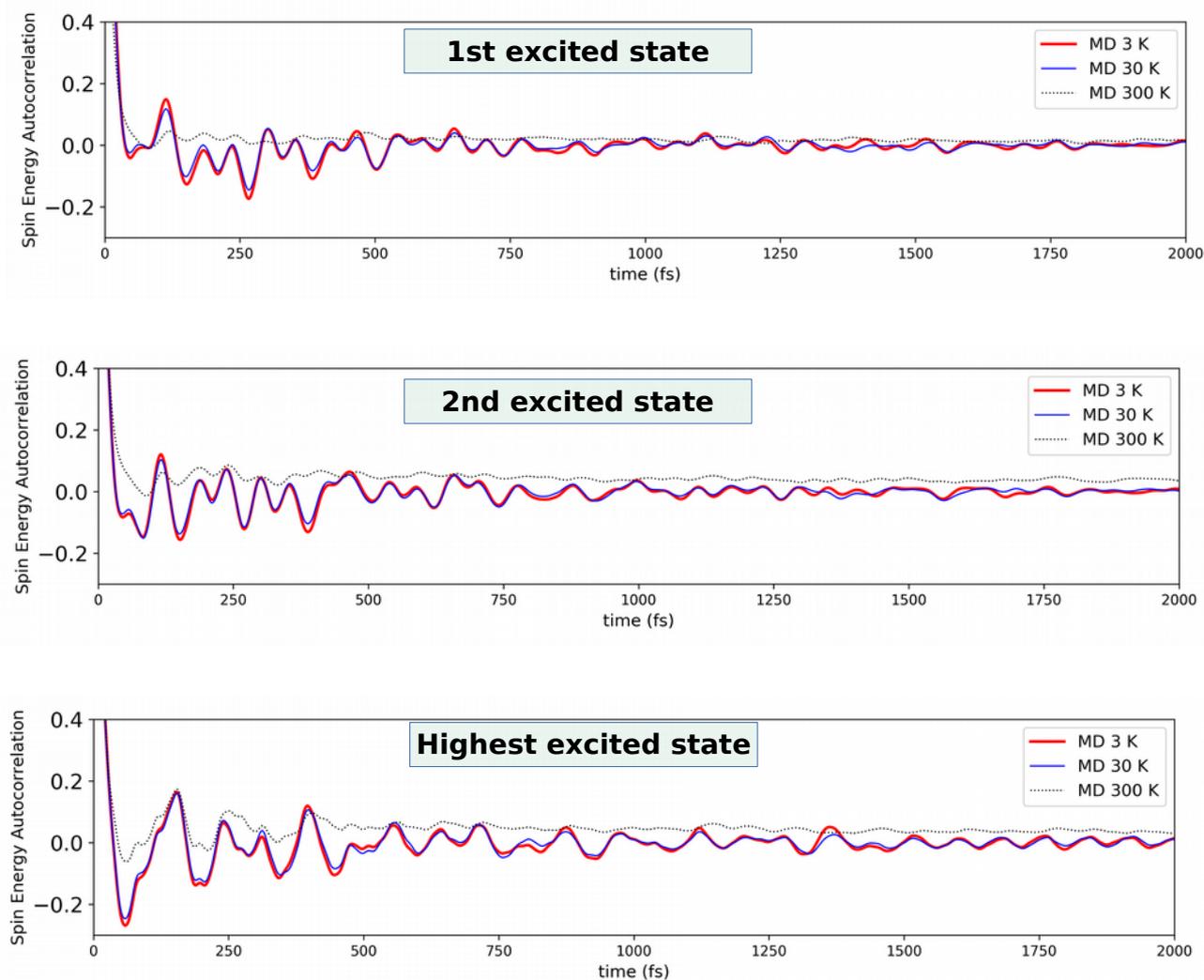

**Figure S5.** Autocorrelation plots of the spin energy levels for the first, second and highest excited spin states of the ground $J$ multiplet resulting from the MD simulations at 3 K, 30 K, and 300K.

## S6. Fast Fourier Transform plots

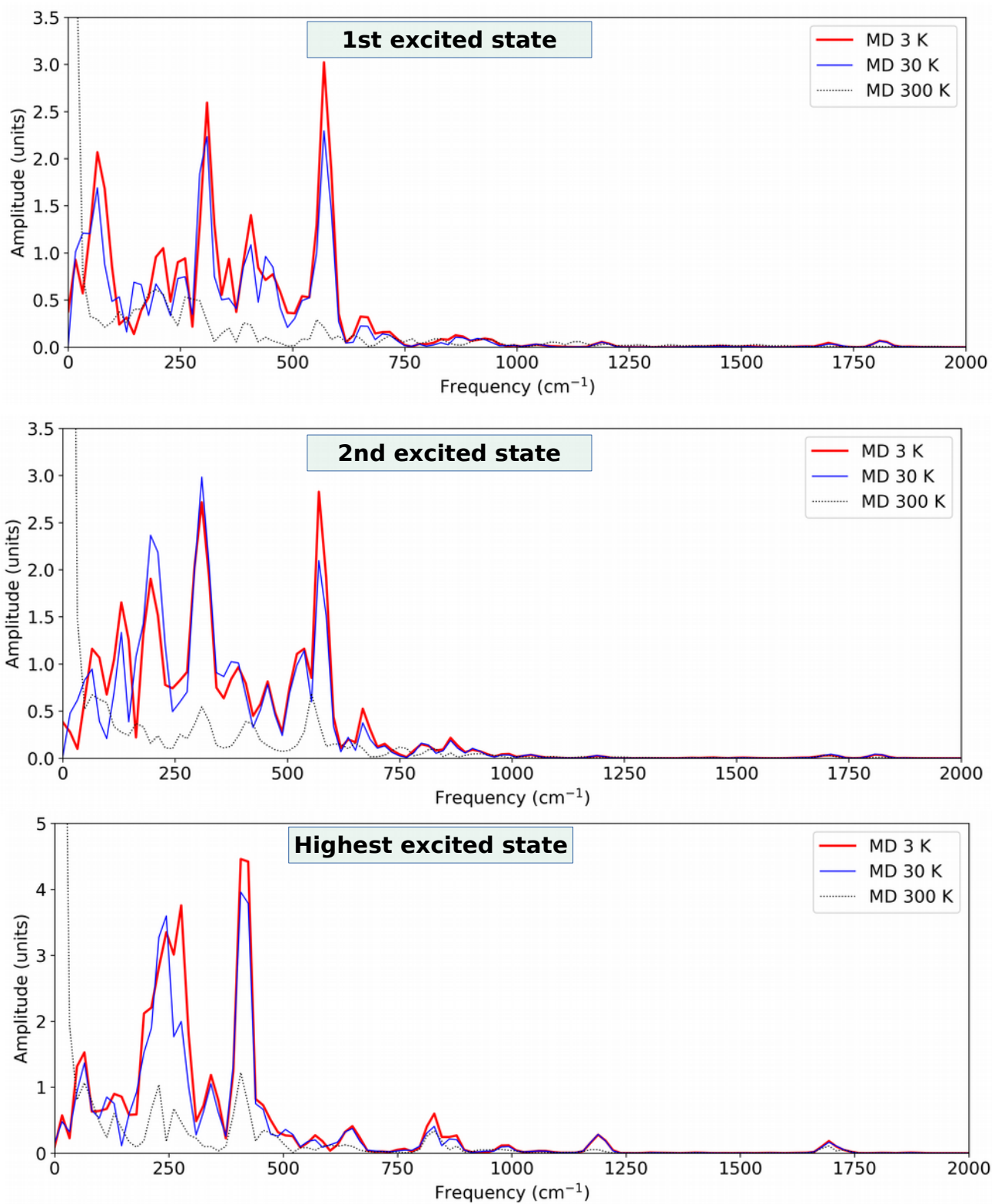

**Figure S6.** Windowed FFT signals resulting from first, second and highest excited states of the ground *J* multiplet for the simulations at 3 K, 30 K and 300 K. The FFT algorithm with a Blackman windowing function has been applied on the autocorrelation of the spin energies (see Fig. S5).